# A Spatial Crypto Technique for Secure Data Transmission

**Sk. Sarif Hassan**
*Applied Statistics unit, Indian Statistical Institute, Kolkata, India*
*Email: sarimif@gmail.com*

**Pabitra Pal Choudhury**
*Applied Statistics unit, Indian Statistical Institute, Kolkata, India*
*Email: pabitrapalchoudhury@gmail.com*

**Sugata Sanyal**
*School of Technology and Computer Science, Tata Institute of Fundamental Research, India*
*Email: sanyals@gmail.com*

**Abstract:** *This paper presents a spatial encryption technique for secured transmission of data in networks. The algorithm is designed to break the ciphered data packets into multiple data which are to be packaged into a spatial template. A secure and efficient mechanism is provided to convey the information that is necessary for obtaining the original data at the receiver-end from its parts in the packets. An authentication code (MAC) is also used to ensure authenticity of every packet.*

**Category:** E.3

## 1. Introduction:

Security of network communications is the most important issue in the world. Information transactions related to banks, credit cards, and government policies are transferred from place to place with the help of network transmission. The high connectivity of the World Wide Web (WWW) has left the world 'open'. Such openness has resulted in various networks being subjected to multifarious attacks from vastly disparate sources, many of which are anonymous and yet to be discovered. This growth of the WWW coupled with progress in the fields of e-commerce and the like has made the security issue even more important [1, 3].

In practice, in a computer network, data is transferred across the nodes in the form of packets of fixed or variable sizes. In usual practice, the implementation is done by some secured algorithms at the application level on the data and the enciphered data is packetized at lower levels (in the OSI architecture) and sent. Any intruder able to obtain all the packets can then obtain the enciphered data by appropriately ordering the data content of each of these packets. Then, efforts can be made to break the secured algorithm used by the sender. In the process of transmission, if it is possible to prevent any information release as to the structure of the data within the packets, an intruder would know neither the nature of the data being transferred nor the ordering of the content from different packets. This is what our algorithm achieves by using a genomical spatial envelope [4]. We have used genomic steganography in enhancing the security in the algorithm.

## 2. The Algorithm

(A) We use the concept of Message Authentication Code (MAC) as suggested in [Rivest, 1998] to authenticate messages. For a packet of data, the MAC is calculated as a function of the data contents, the packet sequence number and a secret key known only to the sender and the receiver, and then it is appended to the packet. On receiving a packet, the receiver first computes the MAC using the appropriate parameters, and then performs a check with the MAC attached to the packet. If there is no match, then the receiver knows that the packet has been tampered with. A detailed explanation is provided in the next section.

Let the size of an encrypted packet data in a network be denoted as *PS*. *PS* has a value of 1024- bits or 4096- bits typically. The packet data is now to break arbitrarily in N parts. These N parts are to be inserted to into the blank envelope.

The size of total data, number N and the nucleotide sequences with which alignment is to be performed using ClastalW is communicated between the receiver and sender using a secure channel. The packet data size is represented in number of bits required to represent the size of total data sent .We use (N-1)*2 bits (or N-1 ATGC's) to represent packet data number.

## Sender encryption technique: S (P)

1. *Transform the data P to a cipher data C by a crypto transform.*
2. *Now using the transformation $\sigma$ (defined as $\sigma(00)$=A, $\sigma(01)$=c, $\sigma(10)$=G AND $\sigma(11)$=T), transform S to its A,T, C and G texture data E.*
3. *Tear the data E into N packets of arbitrary size where $E_{ijk}$ and $E_{ij(k+1)}$ give together $E_{ij}$ Similarly $E_{ij}$ and $E_{i(j+1)}$ give together $E_i$.*
4. *Size (size of data plus size of packet number) of each packet (which has also undergone all the above transformations) is inserted at the starting of each packet before the packet number.*
5. *Now take two or three (sender choice) nucleotide sequences and perform alignment using ClastalW.*
6. *Wherever nucleotides are different for those two or three sequences in the alignment we replace that particular position by one (ATCG text) from the $E_{ijk...}$until it is exhausted.*
7. *After $E_{ijk...}$is exhausted we insert packet number at next positions where nucleotides are different.*
8. *The remaining positions (if any) where nucleotides are different are replaced randomly by ATGC sequence.*
9. *Find MAC corresponding to the packet data DEij...*
10. *Repeat Step 5, 6, 7 and 8 for each $E_{ijk...}$ for all N parts. Each of these new parts is renamed as $DE_{ijk...}$.*
11. *Send each $DE_{ijk...}$ to the receiver in different time and different networks.*

## Receiver's decryption technique: R (DE)

In receiver end the following steps are required to decrypt the message:

1. *Receive each $DE_{ijk...}$ from the sender in different time and different networks.*
2. *Perform a check on MAC corresponding to each packet. If satisfied proceed to next step.*
3. *Now take the nucleotide sequences (chosen and communicated by sender) and perform alignment using ClastalW.*
4. *Wherever nucleotides are different for those two or three sequences in the alignment we extract (ATCG text) from each DE at that particular position.*
5. *Size of each packet (expressed in fixed number of ATGC's) is extracted from the starting of each packet.*
6. *Use the transformation $\sigma'$ (defined as $\sigma'(A)$=00, $\sigma'(C)$=01, $\sigma'(G)$=10 and $\sigma'(T)$=11) on the data size.*
7. *Conserve the data packet up to the determined data size and reject the remaining ATGC's and rename the packet as E.*
12. *The packet number is extracted using $\sigma'$ from the end of each packet and the packets are ordered and joined according to packet numbers i.e. $E_{ijk}$ and $E_{ij(k+1)}$ give together $E_{ij}$ Similarly $E_{ij}$ and $E_{i(j+1)}$ give together $E_i$.*
8. *Now using the transformation $\sigma'$ convert the A,T, C and G texture data into cipher text C.*
9. *Transform the cipher data C to a data P by a crypto transform.*

**(B) Demonstration of the Algorithm:**

The plain text (P) is

I AM SUGATA SANYAL

The binary text (B) corresponding to (P):

010010010010000001000001010011010010000001010011010101010100011101000001010101000100000100100 0000101001101000001010011100101100101000001010011100  (Size is 144)

The cipher text (C) corresponding to (B) The crypto map is T (0) =1; T (1) =0.

1011011011011111101111101011001011011111101011001010101011100010111110101010111011111011011 111101011001011110101100011010011010111110101110011

We have to transfer the each data into its ATCG from by the following transformation: $\sigma(00) = A;\ \sigma(01) = C, \sigma(10) = G\ and\ \sigma(11) = T$

The transformed text (E) corresponding to (C) is:

GTCGTCTTGTTGGTAGTCTTGGTAGGGGGTGAGTTGGGGTGTTGTCTTGGTAGTTGGTACGGCGGTTGG TAT

Now we need to break the data E into arbitrarily N parts.

**$E_{1=000001}$**: GTCGTCTTGTTGGTAGTCTTGGT

(Size 23)

$E_2$:AGGGGGTGAGTTGGGGTGTTGTCTTGGTAGTTGGTACGGCGGTTGGTAT (Size 49)

**$E_{21=001001}$**: AGGGGGTG (Size 8)

**$E_{22}$:** AGTTGGGGTGTTGTCTTGGTAGTTGGTACGGCGGTTGGTAT(Size 41)

**$E_{221=101001}$**: AGTTGGGGT (Size 9)

**$E_{222=101010}$**: GTTGTCTTGGTAGTTGGTACGGCGGTTGGTAT (Size 32)

Suppose we take the break data set as { $E_1$, $E_{21}$, $E_{221}$, $E_{222}$ }

Now we add packet size (size of data and size of packet number) at the starting of each packet.

Since our data size before packet formation is 144 which can be represented in (10010000) 8 bits we use a 8 bits (4 ATGC's) representation of packet size.

Sizes of each packet are given below:

$E_1$: 26 (size of packet + size of packet number=23+3)
  $(26)_{10}=(00011010)_2$

Using crypto map we get size as 11100101
Using $\sigma$ size is represented as TGCC

$E_{21}$: 11 (size of packet + size of packet number=8+3)
$(11)_{10}= (00001011)_2$
Using crypto map we get size as 11110100
Using $\sigma$ size is represented as TTCA

$E_{221}$: 12 (size of packet + size of packet number=9+3)
$(12)_{10}= (00001100)_2$
Using crypto map we get size as 11110011
Using $\sigma$ size is represented as TTAT

$E_{222}$: 35 (size of packet + size of packet number=32+3)
$(35)_{10}= (00100011)_2$
Using crypto map we get size as 11011100
Using $\sigma$ size is represented as TCTA

Now we need to packetize the above data into star model (OR1D2, OR1D4 and OR1D5) ….
Note that we need to send the packet number along with this data packet. We insert the data content followed by the packet number.

Therefore the data packets are as follows:

**DE$_{000001}$**: (TGCC) GTCGTCTTGTTGGTAGTCTTGGT (AAC)

ATGGATGGAGTGAACCAGAGTGACCGTTCACAGTTCCTTCTCCTGGGGAT

GTCAGAGAGTCCTGAGCAGCAGCTGATCCTGTTTTGGATGTTCCTGTCCA

TGTACCTGGTCACGGTGCTGGGAAATGTGCTCATCATCCTGGCCATCAGC

TCTGATTCCCTCCTGCACACCCCCTTGTACTTCTTCCTGGCCAACCTCTC

CTTCACTGACCTCTTCTTTGTCACCAACACAATCCCCAAGATGCTGGTGA

ACGTCCAGTCCCATAACAAAGCCATCTCCTATGCAGGGTGTCTGACACAG

CTCTACTTCCTGGTCTCCTTGGTGTCCCTGGACAACCTCATCCTGGCGGT

GATGGCGTATGATCGCTATGTGGCCAACTGCTGCCCCCTCCACTAGTCCA

CAGCCATGAGCCCTTTGCTCTGTGTCTTGCTCCTTTCCTTGTGTTGGGAA

CTCTCAGTTCTCTATGGCCTCGTCCACACCTTCCTCGTGACCAGCGTGAC

CTTCTGTGGGACTGGACAAATCCACTACTTCTTCTGTGAGATGTAATTGC

TGCTGTGGATGGCATGTTCCAACAGCCATATTAATCACACAGGGGTGATT

GCCACTGGCTGCTTCATCTTCCTCACACCCTTGGGTTTCATGAACATCTC

CTATGTACGTATTGTCAGACCCATCCTATAAATGCCCTCCGTCTCTAAGA

AATACAAAGCCTTCTCTACCTGTGCCTCCCATTTGGGTGTAGTCTCCCTC

TTATATGGGATGCTTCATATGGTATACCTTGAGCCCCTCCATACCTACTC

GATGAAGGACTCAGTAGCCACAGTGATGTATGCTGTGCTGACACCCATGA

TGAATCCGTTCATCTACAGACTGAGGAACAATGACATGCATGGGGCTCTG

GGAAGACTCCTATGAATACGCTTTAAGAGGCTCATA

**DE$_{21=001001}$**:  (TTCA) AGGGGGTG (AGC)

ATGGATGGAGTTAACCAGAGTGACAAGTCAGAGTTCCTTCTCCTGGGGAT

GTCAGAGAGTCCTGAGCAGCAGCGGATCCTGTTTTGGATGTTCCTGTCCA

TGTACCTGGTCACGGTGGTGGGAAATGTGCTCATCATCCTGGCCATCAGC

TCTGATTCCCTCCTGCACACCCCCGTGTACTTCTTCCTGGCCAACCTCTC

CTTCACTGACCTCTTCTTTGTCACCAACACAATCCCCAAGATGCTGGTGA

ACATCCAGTCCCAGAACAAAGCCATCTCCTATGCAGGGTGTCTGACACAG

CTCTACTTCCTGGTCTCCTTGGTGCCCCTGGACAACCTCATCCTGGCAGT

GATGGCTTATGAGCGCTATGTGGCCACCTGCTGCCCCCTCCACTAATGCA

CAGCCATGAGCCCTAGGCTCTGTTTCTTCCTCCTATCCTTGTGTTGGGCT

CTGTCAGTTCTCTATGGCCTCCTGCACACCATCCTCTTGACCAGGGTGAC

CTTCTGTGGGACGTGATAAATCCACTACATCTTCTGTGAGATGTACCTAT

TGCTGAGGTTGGCATGTTCCAACAGCCACATTAGTCACACAGAGGTGATT

GCCACGGGCTGCTTCATCTTCCTCAGACCCTTCGGTTTCATGAACATCTC

CTATGTACGTATTGTCAGAGCCATCCTCATAATACCCTCAGTCTCTAAGA

AATACAAAACCTTCTCTACCTGTGCCTCCCATTTGGGTGGGGTCTCCCTC

TTATATGGGAAACTTGGTATGGTCTACCTACAGCCCCTCCATACCTACTC

AATGAAGGACTCAGTAGCCACAGTGATGTATGCTGTGCTGACACCAATGA

TGAAACCTTTCATCTACAGGCTGAGGAACAACGACATGCATGGGGCTCAG

GGAAGAGTCCTAATAAAACGCTTTCAGAGGCTTAAA

**DE$_{221=101001}$**: (TTAT) AGTTGGGGT (GGC)

ATGGATGGAGTTAACCAGAGTGAATAGTCATAGTTCCTTCTCCTGGGGAT

TTCAGAGAGTCCTGAGCAGCAGCGGATCCTGTTTTGGATGTTCCTGTCCA

TGTACCTGGTCACGGTGGTGGGAAATGTGCTCATCATCCTGGCCATCAGC

TCTGATTCCCGCCTGCACACCCCCGTGTACTTCTTCCTGGCCAACCTCTC

CTTCACTGACCTCTTCTTTGTCACCAACACAATCCCCAAGATGCTGGTGA

ACTTCCAGTCCCAGAACAAAGCCATCTCCTATGCAGGGTGTCTGACACAG

CTCTACTTCCTGGTCTCCTTGGTGGCCCTGGACAACCTCATCCTGGCCGT

GATGGCATATGATCGCTATGTGGCCAGCTGCTGCCCCCTCCACTAATGCA

CAGCCATGAGCCCTATGCTCTGTGTCTTCCTCCTATCCTTGTGTTGGGTG

CTATCTGTGCTCTATGGCCTCCTACTCACCGTCCTCCTGACCAGAGTGAC

CTTCTGTGGGACTGGACAAATCCACTACTTCTTCTGTGAGATGTACCTCA

TGCTGAGGTTGGCATGTTCCAACAACCAAATAATTCACACAGAGTTGATT

GCCACAGGCTGCTTCATCTTCCTCATGCCCTTCGGATTCTTGAGCACATC

CTATGTACGTATTGTCAGACCCATCCTATGAATCCCCTCAGTCTCTAAGA

AATACAAAACCTTCTCTACCTGTGCCTCCCATTTGGGTGGCGTCTCCCTC

TTATATGGGATGCTTATTATGGTGTACCTCAAGCCCCTCCATACCTACTC

TATGAAGGACTCAGTAGCCACAGTGATGTATGCTGTGGTGACACCTATGA

TGAAACCGTTCATCTACAGGCTGAGGAACAATGACATGCATGGGGCTCTG

GGAAGAATCCTATGCAAACCCTTTTAGAGGCAAATA

**DE$_{222=101010}$:** (TCTA) GTTGTCTTGGTAGTTGGTACGGCGGTTGGTAT (GGG)

ATGGATGGAGTCAACCAGAGTGATAGTTCATAGTTCCTTCTCCTGGGGAT

GTCAGAGAGTCCTGAGCAGCAGCTGATCCTGTTTTGGATGTTCCTGTCCA

TGTACCTGGTCACGGTGCTGGGAAATGTGCTCATCATCCTGGCCATCAGC

TCTGATTCCCTCCTGCACACCCCCTTGTACTTCTTCCTGGCCAACCTCTC

CTTCACTGACCTCTTCTTTGTCACCAACACAATCCCCAAGATGCTGGTGA

ACGTCCAGTCCCAGAACAAAGCCATCTCCTATGCAGGGTGTCTGACACAG

CTCTACTTCCTGGTCTCCTTGGTGTCCCTGGACAACCTCATCCTGGCAGT

GATGGCGTATGATCGCTATGTGGCCATCTGCTGCCCCCTCCACTAGGTCA

CAGCCATGAGCCCTACGCTCTGTGTCTTGCTCCTCTCCTTGTGTTGGGGG

CTTTCTGTGCTCTATGGCCTCGTTCACACCTTCCTCGTGACCAGGGTGAC

CTTCTGTGGGGCATGAGACATCCACTACATCTTCTGTGATATGTAGCTCA

TGCTGAGGTTGGCATGTTCCAACAGCCAAATTATTCACACAGCGCTGATT

GCCACCGGCTGCTTCATCTTCCTCATGCCCTTAGGTTTCATGATCAGCTC

CTATGTACGTATTGTCAGACCCATCCTTCAAATCCCCTCAGTCTCTAAGA

AATACAAAACCTTCTCCACCTGTGCCTCCCATTTGGGTGTAGTCTCCCTC

TTATATGGGAGTCTTCTTATGGTATACCTAGAGCCCCTCCATACCTACTC

ATTGAAGGACTCAGTAGCCACAGTGATGTATGCTGTGCTGACACCAATGA

TGAAACCCTTCATCTACAGGCTGAGGAACAAAGACATGCATGGGGCTCTG

GGAAGATTCCTATACAAACCCTTTAAGAGGCCAATA

The above packets are sent at various times and different networks. The data packet size can be considered as the header and packet number as the trailer of data packet. We consider that two information the total message length and number of packets N (144 and 4 respectively for above example) has been communicated to the receiver through a secure channel. These packets when received on the receiver's end can be decrypted using the keys (nucleotide sequences used in ClustalW, the crypto transform and the relation $\sigma$ ) and following the steps mentioned in the algorithm.

## 3. Comments on the security aspects of the proposed algorithm

In the proposed algorithm, the data, ready to be sent, is broken into N arbitrary parts. These parts are to be sent through different communication channels at different times. As a consequence, it would be almost impossible for any attacker to get back all those parts. Even if one gets all the parts again it would be difficult for him/her to get the original information by properly merging the parts because key packets have to be recognized properly.

In our algorithm a steganographic method is used, the strength of DNA steganography lies in our conjecture that deciphering on the basis of cryptanalysis techniques like chosen plain text attack, Chosen-cipher text attack or adaptive chosen-cipher text attack etc become fruitless because it is mathematically infeasible to extract the whole information from these parts in reasonable time.

## 4. Concluding remarks and future endeavors:

The aim of the any cryptosystem is to procure a reasonable amount of security in transmission of a packet data. Although our algorithm is not in the form of conventional cryptosystem due to the fact that in our system there is a provision of key packet sending but definitely our system deserves the attention of the crypto-community in terms of the security of transmission of data.

In spite of our best efforts we have not been successful in deciphering any appreciable information encoded in the DNA sequences for the last decades. So we are hopeful that we will be able to settle the conjecture as posed in the section 3 that it would be computationally infeasible by any present cryptanalytic techniques to decipher the encoded information.

## Acknowledgements

Authors are grateful to **Dr. Rangarajan Athi Vasudevan** for his valuable suggestions and visiting students **Mr. Rajneesh Singh, Ms. Snigdha Das** and for their valuable technical help in making advanced C programs and other computer applications on Windows support used for this study.

## References:

1. Vasudevan R. A., Abraham A and Sanyal S., "*A Novel Scheme for Secured Data Transfer over Computer Networks*", Journal of Universal Computer Science , Vol 11, Issue 1, pp 104-121, 2005

2. Ashish Gehani, Thomas LaBean, and John Reif., "*DNA-Based Cryptography*", DIMACS DNA Based Computers V, American Mathematical Society, 2000.

3.  R. A. Vasudevan, A. Abraham, S. Sanyal and D. P. Agrawal, "*Jigsaw-based Secure Data Transfer over Computer Networks*", IEEE International Conference on Information Technology: Coding and Computing, 2004. (ITCC '04), Proceedings of ITCC 2004, Vol 1, pp 2-6, April, 2004, Las Vegas, Nevada.

4.  Sk. S Hassan, P Pal Choudhury, A Pal, R. L. Brahmachary and A. Goswami L-Systems: A Mathematical Paradigm for Designing Full Length Genes and Genomes, (2010) (Communicated to Journal).